**Author's name**
Soheila Khajoui, Saeid Dehyadegari, Sayyed Abdolmajid Jalaee

**Affiliations**
- Department of Management, Faculty of Management & Economics, Shahid Bahonar University of Kerman, Kerman, Iran
- Department of Management, Faculty of Management & Economics, Shahid Bahonar University of Kerman, Kerman
- Department of Economics, Faculty of Management & Economics, Shahid Bahonar University of Kerman, Kerman, Iran

**Author's email address**

- s.khajoui@aem.uk.ac.ir

- dehyadegari@uk.ac.ir

- jalaee@uk.ac.ir

**Corresponding author:**

Saeid Dehyadegari:

- Department of Management, Faculty of Management & Economics, Shahid Bahonar University of Kerman, Kerman, Iran
- dehyadegari@uk.ac.ir

# Predicting the impact of e-commerce indices on international trade in Iran and other selected members of the Organization for Economic Co-operation and Development (OECD) by using the artificial intelligence and P-VAR model

**Abstract**
This study aims at predicting the impact of e-commerce indicators on international trade of the selected OECD countries and Iran, by using the artificial intelligence approach and P-VAR. According to the nature of export, import, GDP, and ICT functions, and the characteristics of nonlinearity, this analysis is performed by using the MPL neural network. The export, import, GDP, and ICT findings were examined with 99 percent accuracy. Using the P-VAR model in the Eviews software, the initial database and predicted data were applied to estimate the impact of e-commerce on international trade. The findings from analyzing the data show that there is a bilateral correlation between e-commerce which means that ICT and international trade affect each other and the Goodness of fit of the studied model is confirmed.

**Introduction**
Today, one of the debates on which the economy perfectly depends is online space. A significant part of the Gross Domestic Production (GDP) generates in the online environment, and this is the online space that could contribute to the dynamic of today's economics in countries [1]. In other words, the online environment could significantly decrease the costs of performing economic actions in easing business relationships and reducing time and resources, and this justifies the important role of the online space in the dynamic of countries [2]. Technological advances affect international trade and cause improvement and development of international trade so that millions of people take advantage of the Internet for different purposes including research, and online shopping of products [3]. Prediction is a quantitative assessment of the likelihood of future events occurring that is calculated regarding the previous and present information. The previous and present information is presented in a series of equations or time series patterns and future events are predicted by using the next series of patterns and information. Several techniques exist for modeling and predicting time series. The statistically traditional ways including moving average, weighted average, and ARIMA, linearly predict future amounts of variables. For several past decades, linear methods have been primarily used in scientific studies, due to the simplicity of these approaches in comprehension and implementation. Despite some advantages, the linear models suffer from disadvantages including a lack of ability in predicting the nonlinear associations. To address the deficiencies in linear techniques, nonlinear methods such as artificial intelligence which possess the advantage of the flexibility to predict various nonlinear models, have been recently applied in time series predictions [4].

Furthermore, researchers of different branches of science have focused their major studies on predicting the variables and the studied phenomena. There exists no exception among the economic and commercial researchers, and we can readily realize the importance of prediction if the various prediction techniques are reviewed, including artificial intelligence. Predicting the trade contributes to governments and companies crafting policies and making more efficient decisions, assessing the production and readiness to perform international trade in the future [5]. Hence, trade prediction plays a key role in the organization to assess the correlation between internal decision-making and external factors that are out of control. Traditional time series techniques operate based on the trends existing among the data and they fail to predict macroeconomic changes in trade and to discover nonlinear and advanced patterns but approaches of artificial intelligence and neural networks are completely efficient and absorb attention in the prediction of nonlinear, advanced patterns and big data [6]. Artificial intelligence is one of the significant technologies mostly used in e-commerce; today, the technology of artificial intelligence has the greatest impact on e-commerce and this effect will increase in the future. By analyzing the consumption pattern and predicting present needs, artificial intelligence contributes to operating trade systems to predict future demand. When the efficiency increases, costs will decrease and the trading performance will improve. Therefore, artificial intelligence plays a role in international trade as one of the e-commerce achievements, and today the world is moving to artificial intelligence. Since many economic firms consider international trade and e-commerce major elements, it is required to pay special attention to these factors, to save time, costs, and challenges that a firm may encounter in the trading area, in the future. As e-commerce indicators are considered novel, then, it is important to have a compounded sample of three groups of developed, new developed, and developing countries. Hence, the present research focuses on the selected OECD countries and Iran that create a sample of countries with those characteristics mentioned previously. A group of these countries is Iran's trade partners and another group is potential trade partners which have a positive impact on trade strategies. As a result, as discussed, this study aims at predicting the effect of e-commerce indicators on international trade among the selected OECD countries and Iran by using the artificial intelligence approach. The study is organized as follows: in the first section, we present the literature review. Section 2 discusses the research background. Section 3 presents the methodology and research framework. Section 4 provides the results, data, and model and finally, section 5 presents the conclusion and discussion.

**Literature Review**
With the emergence of the Internet, e-commerce has been considered important by societies. E-commerce has become a channel for purchasing for people [7] and it is found a novel driver in the retailing industry [8]. The high effect of the Internet has paved the path for the success of different businesses in e-commerce including eBay, Yahoo, Google, and PayPal [9]. International trade has been one of the key characteristics of human-beings societies for ages [10] and multinational companies involved in the global supply chain are increasingly opening up to international trade [11]. International trade not only covers goods exchange but also requires a permanent exchange of information, values, and attitudes, considering all factors, and these affect the preferences of representatives. In addition to economic issues, international trade has direct consequences on the preferences and interests of representatives. International trade consequences have been developed in the emerging economic literature about the preferred factors in the relationship between globalization and cultural diversity. By using different models, the concept of trade has been reviewed in relationship with other concepts, including cultural values, consumption customs, and preferences which shows the cultural interaction after the trade and its consequences on the local culture of each country. Representatives not only compare their consumption with a local representative but also with the representatives of trade partner countries [12]. Also, the trend of international trade developments in the future, including e-commerce, recent trade ways, and major infrastructure development, will lead to pressure on national borders that soon excels the available resources to intervene. In fact, e-commerce facilitates delivering goods and services to countries and throughout the world [10]; therefore, e-commerce is considered a dominant direction in the business environment [13]. In addition, e-commerce causes economic growth and international trade in OECD countries. As a result, to benefit from complete information technology potential, training aspects should be considered in the policy-making and planning for economic growth [14].

According to the linearity and nonlinearity of time series data, the predictive time series models, such as ARIMA are efficient in linear associations, while predictive models based on artificial intelligence and neural networks are appropriate for nonlinear databases, including data concerning trade [15]. Considering its definition, history, types, characteristics, and financial relations, artificial intelligence is mostly discussed in the financial markets of trade. Artificial intelligence is a complex of technologies that combine computers and algorithms to simulate and strengthen human intelligence. Artificial intelligence is an optimization machine that enables comparative patterns to discover and make the best prediction by using data and statistical frequency techniques [16]. Artificial neural networks (ANN) are neural networks for data processes inspired by the biological neural system and can be used as a predictive model of specific software when enough training is received [17].

**Research Background**
 predictive models for financial time series are mostly considered in trade and science discussions, and neural networks perform accurately, regarding machine learning algorithms and discovering patterns of big data (such as time series). Moreover, combined neural network models, including WDNN, and WNN, perform more efficiently in the financial market and long-term prediction [18].

Today inclusive development of e-commerce has been counted as a key motive of economic growth in different societies. As a result, considerable and increased effects of information and communication technology cause producing distinctive words and equivalents in the scientific literature, such as information economics, networking economics, startup economics, digital economics, and sharing economics [19].

E-commerce prosperity brings about a considerable amount of data on bought items online. However, the majority of data is redundant and decreases the user recovery efficiency. This not only reduces user satisfaction but also imposes huge loads on the websites of e-commerce; hence, researchers of e-commerce pay great attention to efficiently and accurately finding ways that discover users from favorable big data. The personalized recommendation system enables delivering appropriate items to users, regarding purchase background, review, and rates, and as a result, reduces their access time to such features [7].

International trade is mostly considered a play that is presented by national governments; therefore, changes in international trade patterns are realized as a result of technical process change occurring due to international technical knowledge sharing, as well. In fact, the key players are multinational companies or large OEMs taking place on the top of the global supply chains. Trade is an integral part of the strategies of cost minimization [3].

The findings indicate that SMEs without e-commerce development have more income and employment growth compared to their peers in other industries. Benefiting from electronic information exchange will decrease the cost of production, process, and use of information. Electronic exchange, also, removes geographical distances and time differences among markets, and international trade is establishing an integrated and borderless platform. In other words, recent factors shaping the modern economy, including information, information technology, regulation, globalization phenomenon, the Internet, and management insights can facilitate economic activities [20].

E-commerce affects the economy and export development. Hence, the customers can buy cheaper goods and sellers earn more profits due to the mediator removal and other costs such as location cost. Meanwhile, virtual businesses are created in the e-commerce platform enabling individuals to take several jobs, and individuals will earn money from the electronic business. Contribution to international trade development is the other advantage of e-commerce that leads to addressing traditional trade barriers. For example, the aspect of distance which was one of the great concerns of merchants and put goods in danger can be removed by using e-commerce, it accelerates transactions, and by taking advantage of the Internet and e-commerce can create a comparative advantage [21].

Internet improves international trade, while it removes trade barriers; therefore, international trade volume will increase through e-commerce. OECD countries with high GDP based on their imports can benefit from the above knowledge. Also, e-commerce can significantly impact the service trade. In addition, it is expected that e-commerce, directly and indirectly, creates or eliminates jobs. New jobs will develop in the field of information technology and communication, however, increasing demand and productivity will lead to indirect job creation. Net profit and loss of employment depend on demand for specific skills [22].

**Method**

This research is considered a national future study and is practical in terms of purpose which predicts the data on export, import, GDP, and ICT variables by using an MLP neural network for 5 years (from 2021 to 2025) and estimates the impact of ICT on international trade by using P-VAR model, in Iran and other selected OECD countries including the United States, Canada, Germany, France, Japan, Turkey, South Korea, Portugal, and Greece. The data on variables including export, import, GDP, and ICT, in selected countries, has been collected through reliable websites such as the World Bank, then, the amount of export, import, GDP, and ICT is accordingly predicted by MLP neural network in Pycharm software, from 2021 to 2025. The real dataset and predicted data were used to predict the effect of ICT on international trade by using the P-VAR model in E-views software. MPL neural network and P-VAR model are explained as follows.

**Artificial Neural Network**

Artificial Neural Network (ANN) is a well-organized data mining technique and operates based on biological neural networks. ANN collects a huge amount of connected data in particular patterns joined through units called nodes or neurons, and each of these neurons has a mutual connection with other neurons. Each connection has a specific weight sending feedback about input data. This part plays an essential role in neurons in terms of arising a particular problem. Each neuron remains in a synthetic state or keeps an activation signal. Output signals are produced as they have joined to input signals and activation functions are sent to other units [23]. Figure 1.

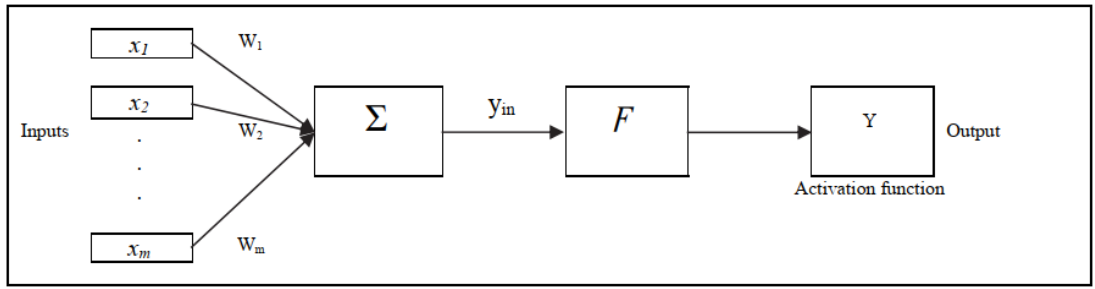

Figure1- Artificial Neural Network [23]

**MLP Neural Network**
Multilayer perceptron neural network is the most common neural network. These networks are considered Feedword Neural Networks that can accurately provide a nonlinear mapping by choosing efficient layers and neurons. Controllable parameters in the MLP networks are the weight of connections among layers and the process of learning is identifying appropriate values for the weights of connections among neurons. The most popular learning algorithm in these networks is the backward propagation of errors. In feedforward neural networks, neurons are arranged in the layers in which the beginning layer (input) starts and ends in the output layer. Also, there is Hidden Space in this network, and each layer has one or more neurons. Neurons of one layer join to other layers and the output of a neuron of a layer only depends on connected neurons of previous layers and the weights of connections. The major part of optimization in the multilayer perceptron structure is to determine the number of hidden layers and neurons of each hidden layer to approach the least errors. A proposition exists in the theories of artificial neural networks that suggests a hidden layer with a sufficient number of neurons enables the estimation of any nonlinear relationships. Figure 2 simply illustrates the structure of a multilayer perceptron neural network in which the input layer, hidden layer, and output layer are highlighted.

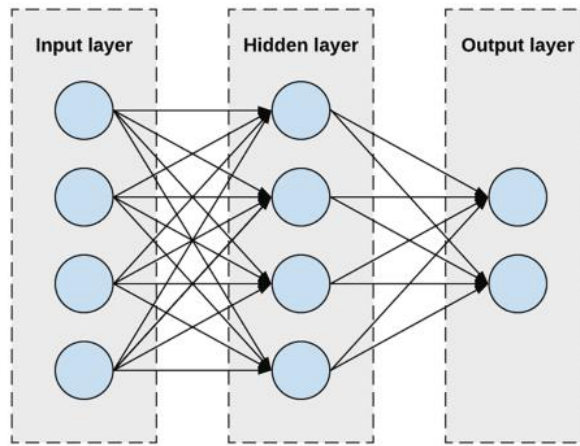

Figure 2- The structure of multilayer perceptron artificial neural network [24].

There exist connections among neurons of distinctive layers having a specific weight. Through the learning process, these weights and added constant values called bias are repeatedly changing to estimate the errors and to minimize the real values. Activation functions are applied to transfer the output layers of each layer to other layers. There are different kinds of activation functions, including linear function, Sigmoid function, hyperbolic tangent function, etc.

## P-VAR Model

This study applies the Panel Var model to estimate the coefficients of variables and to extract the stimulus-response function. The Panel Var model is presented as follows:

$$X_{it}= \Gamma L \, X_{it}+U_i + \epsilon_{it} \qquad (1)$$

Where in equation (1), $X_{it}$ is the dependent variable, $\Gamma L$ is the polynomials matrix of lagged dependent variable that is presented as follows $\Gamma L = \Gamma_1 L_1 + \Gamma_2 L_2 + \cdots$, Ui is the fixed effects vector and $\epsilon_{it}$ is the special error vector.

Data analysis

To predict export, import, GDP, and ICT, the research uses an MLP neural network and the value of export, import, GDP, and ICT is predicted with a 99% accuracy rate, for 5 years (20 quarters), from 2021 to 2025, in Iran and the selected OECD countries including the United States, Canada, France, Germany, Japan, Turkey, South Korea, Portugal, and Greece. Then, by using the P-Var model, the impact of e-commerce on international trade is assessed, regarding the real data, from 2000 to 2020 and the predicted data on variables, in E-views. In addition, the applied model in the research is as follows:

$$(XP+MP)/GDP=f\,(ICT) \qquad (2)$$

XP and MP are export and import, respectively, GDP is the dependent variable, and ICT is the independent variable.

The findings from the P-Var model are presented as follows:

## Stationary Test for independent and dependent variables

Testing for stationary of all variables involved in the examination is a prerequisite to examining the model because using non-stationary time series or panel data in the regression produces spurious results. To test the stationary of data, ADF - Fisher Chi-square ،PP - Fisher Chi-square ،Levin, Lin & Chu t are used, and Tables 1 and 2 provide the results for all variables. If the calculated statistic is greater than the common value in the confidence interval, and also the statistic is less than 0.05, then, the null hypothesis based on the non-stationary data is rejected. The results provided in Tables 1 and 2 alongside checking values of statistics and the possibility of acceptance show that the null hypothesis based on non-stationary data is rejected, meaning that all variables are stationary except the non-stationary dependent variable that by transforming to a level of differencing become stationary data.

Table 1- Stationary Test for dependent variables (XP+MP/GDP)

| Method | Statistic | Prob.** | Cross-sections | Obs |
|---|---|---|---|---|
| Null: Unit root (assumes common unit root process) | | | | |
| Levin, Lin & Chu t* | -7.5265 | 0.0000 | 10 | 900 |
| Null: Unit root (assumes individual unit root process) | | | | |
| ADF - Fisher Chi-square | 107.359 | 0.0000 | 10 | 900 |
| PP - Fisher Chi-square | 175.873 | 0.0000 | 10 | 980 |

Table2- Stationary Test for independent variables (ICT)

| Method | Statistic | Prob.** | Cross-sections | Obs |
|---|---|---|---|---|
| Null: Unit root (assumes common unit root process) | | | | |
| Levin, Lin & Chu t* | -4.28222 | 0.0000 | 10 | 896 |
| Null: Unit root (assumes individual unit root process) | | | | |
| ADF - Fisher Chi-square | 44.487 | 0.0013 | 10 | 896 |
| PP - Fisher Chi-square | 51.6539 | 0.0001 | 10 | 990 |

**P-Var Model Examination**

In this section, we assess the effect of variables in the model, and the results from the examination are presented in Table 3:

Table3- The effect of variables in the model

| Vector Autoregression Estimates | | |
|---|---|---|
| Sample (adjusted): 2003Q3 2025Q4 | | |
| Included observations: 900 after adjustments | | |
| Standard errors in () & t-statistics in [ ] | | |
| | F | ICT |
| F | -1.615135 | 0.657734 |
| | (0.17869) | (0.30114) |
| | [-9.03873] | [2.18412] |
| ICT | 0.044439 | 0.404303 |
| | (0.04458) | (0.07513) |
| | [0.99682] | [ 5.38133] |

|   |   | -0.004523 | 0.073125 |
|---|---|---|---|
|   | C | (0.00965) | (0.01626) |
|   |   | [ -0.46893] | [ 4.49843] |
| R-squared |   | 0.973314 | 0.860430 |
| F-statistic |   | 8160.766 | 1379.383 |

(Which F is the independent variable of the research, i.e., $XP + MP/GDP$)

According to Table 3, it is realized that variables have interactions. ICT, and (XP+MP/GDP) have significant impacts on each other. According to the examination, international trade has a significant impact on e-commerce with a 0.657734 coefficient, and e-commerce significantly influences international trade with a 0.404303 coefficient, as well. These coefficients show positive and significant impacts of these variables on each other.

**The Analysis of stimulus-response function**
In the Panel-Var model, when the variables are examined, checking the interactions and dynamics of variables is the most important part. Regarding the discussion, the effect of e-commerce (ICT) on international trade (XP+MP/GDP) is assessed in this section.

Figure 3 shows a dynamic response of international trade to shocks of e-commerce. In this figure, the horizontal axis is time while the vertical axis shows the standard deviation.

Figure3- Response of F to ICT innovation

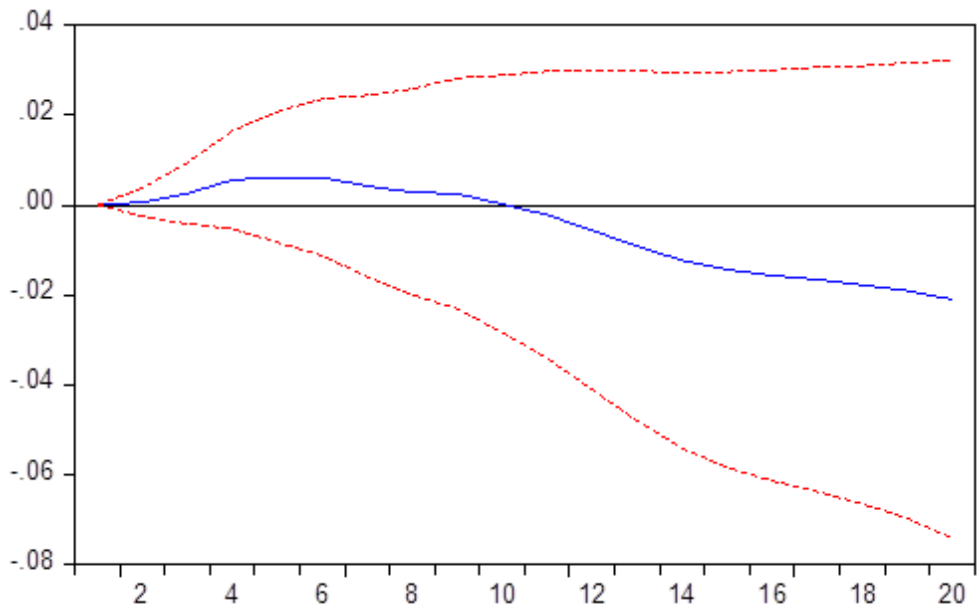

In Figure 3, the stimulus-response function shows that the response of international trade to e-commerce is tangible. On the other hand, when a shock affects e-commerce, it will change international trade, and as a result, the influence of shock is increasing over time.

## Conclusion

Generally, nonlinear models should be used to examine and to choose the examination model, in terms of prediction, due to the nature of export, import, GDP, and ICT functions in developed and underdeveloped countries (generally relationships in this function are inherently nonlinear). Hence, the MLP neural network model has been used to predict export, import, GDP, and ICT in the selected OECD countries and Iran, for the characteristic of flexibility and its potential to model nonlinear relationships. The findings from this model show considerable reliability. Time series data of export, import, GDP, and ICT were collected in the selected countries, from 2000 to 2020; the data is used to predict with a 99% accuracy rate, for the 2021-2025 period. Then, by taking advantage of the P-Var model, the impact of e-commerce on international trade is examined by using the real dataset and predicted data. The results show that first, the research variables are stationary. Second, e-commerce interacts with international trade in Iran and the selected OECD countries including the United States, Canada, Germany, France, Japan, Turkey, South Korea, Portugal, and Greece. Both e-commerce and international trade influence each other.

## Acknowledgments

This research did not receive any specific grant from funding agencies in the public, commercial, or not-for-profit sectors.

## declaration of interest statement

On behalf of all authors, the corresponding author states that there is no conflict of interest.